\documentstyle[epsf,11pt]{article}

\begin{document}

\title{Flux flow resistivity in the system with multiple superconducting gaps\thanks{Submitted to SCES'04}}
\author{Jun Goryo and Hiroshi Matsukawa \\
{\it Department of Physics and Mathematics, Aoyama Gakuin University}, \\
{\it 5-10-1 Fuchinobe, Sagamihara, Kanagawa, 229-8558, Japan}}
\date{}
\maketitle

\begin{abstract}

We calculate the flux flow resistivity in a superconductor 
with multiple $s$-wave superconducting gaps. Our 
result agrees well with anomalous field dependence of the 
resistivity recently observed in the two-gap superconductor MgB$_2$.

\end{abstract}

\begin{flushright}
\end{flushright}

The vortex lines under an applied current catch 
the Lorentz force and move perpendicular to the current 
and magnetic field when the Lorentz force 
becomes larger than the pinning force. 
This is the flux flow state and the finite resistivity arises\cite{Tinkham}.   
The essence of the flux flow resistivity is the presence of 
the bound states inside the vortex core with small energy gap, 
so that  the conductivity in the core is practically 
normal in the usual temperature range.  Such states were found by Caroli, de Gennes 
and Matricon by using the microscopic methods\cite{Tinkham}. 
Bardeen and Stephen found that the flux flow resistivity is basically proportional to the magnetic field $H$ in both of the dirty and moderately clean $s$-wave superconductors\cite{Tinkham,matsuda1}.  
This relation holds well almost all of the field range in the vortex state.  

The superconductivity of MgB$_2$ has the highest transition temperature ($T_c\simeq39K$) 
in the metallic materials at present\cite{nagamatsu} and 
the great amount of investigations has been done\cite{review}.  
As one of the characteristic features in this superconductor, we can point out the fact 
that two different superconducting gaps with $s$-wave pairing
symmetry exist on the two separated Fermi surfaces having roughly  equal density of states (DOS)
\cite{review}. Recently, the measurement of the flux flow resistivity in 
MgB$_2$ was reported by Matsuda's group\cite{matsuda}. 
Large deviations from the $H$-linear dependence for the flux flow 
resistivity has been found, in spite of the fact that the pairing symmetry of 
MgB$_2$ is $s$-wave\cite{review}. 
Such anomalous behavior will be related to the multi-gap 
structure, but a clear explanation has not been shown so far.  
In this paper, we investigate the flux flow state in the multi-gap model and  
propose new mechanism for the anomalous flux flow resistivity. 

First, let us remind the flux flow in the system with single superconducting band with a 
$s$-wave gap\cite{Tinkham}.  We suppose that the magnetic field ${\bf H}=H\hat{\bf z}$  
and that the pinning force is negligible\cite{matsuda}.  
The velocity of flux lines ${\bf v}$ is related to the applied 
current ${\bf J}$ by the balance equation 
\begin{equation}
{\bf J}\times \hat{\bf z}\frac{\Phi_0}{c}=\eta {\bf v},  
\label{balance}
\end{equation}
where $\Phi_0=h c / 2 e$ is the unit flux and the l.h.s. and the r.h.s of Eq. (\ref{balance}) denote the Lorentz force and viscous force, respectively. 
The origin of viscosity $\eta$ is the dissipation 
caused by the Calori-de Gennes-Matricon bound states inside the core 
with small energy gap\cite{Tinkham}. The dissipation induces 
the electric field  and its macroscopic value is ${\bf E}= {\bf H}\times{\bf v} /c$. 
By using this equation and Eq. ({\ref{balance}), we can derive ${\bf J}={\bf E}/\rho_f$, 
where $\rho_f=\Phi_0 H / \eta c^2$ denotes the flux flow resistivity. 
As we mentioned, Bardeen and Stephen found that $\rho_f$ is proportional to the magnetic field, namely, $\rho_f=\rho_n H / H_{c2}$,  where $\rho_n$ is the normal state resistivity\cite{Tinkham}. 
Then, we obtain $\eta$, which is independent of $H$, and ${\bf v}$ is determined by Eq. (\ref{balance}). 

Next, we discuss the multi-gap system. We suppose that 
there are two superconducting bands 
with different-sized $s$-wave gaps in the system. The vortex states appear for each bands 
 when the magnetic field greater than the lower critical value is applied to the system. 
The applied current ${\bf J} $ is divided and the flux flow states 
 arise for each two bands. Let ${\bf J}_1$ and ${\bf J}_2$ be the divided current 
 in 1st-band and that in 2nd-band, respectively.  
 We suppose that vortices in the 1st-band  and that in the 2nd-band have the same velocity, since 
 two kind of the cores would be overlapped  because of the nature of the superconductivity. 
 
 We extend the balance equation Eq. (\ref{balance}) to the two-gap system, that is 
 \begin{equation}
 {\bf J}_i\times \hat{\bf z} \frac{\Phi_0}{c}=\eta_i {\bf v}, 
 \end{equation}
 where $\eta_i$ $(i=1,2)$ is the viscosity for the $i$-th band vortices. 
By using the equation ${\bf E}= {\bf H} \times {\bf v} / c$,  
we can derive 
${\bf J}_i={\bf E}/\rho_f^{(i)}$, 
where $\rho_f^{(i)}=\Phi_0 H / \eta_i c^2$ denotes the flux flow resistivity in $i$-th band. 
 Then, 
 \begin{equation}
{\bf J}=\sum_{i=1}^{2}{\bf J}_i=\sum_{i=1}^{2} \frac{1}{\rho_f^{(i)} } {\bf E},  
\label{ext-j-t-b}
\end{equation}
and this equation indicates that the total flux flow resistivity 
in this system is 
\begin{equation}
\frac{1}{\rho_f^{two}}=\sum_{i=1}^{2}\frac{1}{\rho_f^{(i)}}, 
\label{rho-f-two}
\end{equation}
which reminds us the parallel connection between two different resisters. 

The resistivity $\rho_f^{(i)}$ may obey the Bardeen-Stephen relation, i.e., 
\begin{equation}
\rho_f^{(i)}=\frac{H}{H_{c2}^{(i)}} \rho_n^{(i)},  
\label{ffr-1-2}
\end{equation}
where $H_{c2}^{(i)}$  and $\rho_n^{(i)}$ are 
the upper critical field for the vortex state and normal-state resistivity in  the 
 $i$-th band, respectively.


Let us compare $\rho_f^{two}$ in Eq. (\ref{rho-f-two}) and 
the experimental results. First we would like to check the field dependence 
of $\rho_f^{two}$. Suppose that $H_{c2}^{(1)} > H_{c2}^{(2)}$. 
By using  Eq. (\ref{ffr-1-2}), 
\begin{equation}
\rho_f^{two}
=
\left\{
\begin{array}{ll}
\frac{H}{\frac{H_{c2}^{(1)}}{\rho_n^{(1)} } + \frac{H_{c2}^{(2)}}{\rho_n^{(2)}}} 
& ({\rm lower~critical~field}<H \leq H_{c2}^{(2)}),\\
& \\
\frac{H}{\frac{H_{c2}^{(1)}}{\rho_n^{(1)} } + \frac{H}{\rho_n^{(2)}}} & 
(H_{c2}^{(2)} < H \leq  H_{c2}^{(1)}).  
\end{array}
\right.
\label{field-vs-rho-f-two} 
\end{equation}
Then, $\rho_f^{two}$ is $H$-linear in the low field regime, 
convexly in the high field regime and continuously connected to 
the normal-state resistivity in the two-band system $(1/\rho_n^{(1)} + 1/\rho_n^{(2)})^{-1}$.  

The flow resistivity measurement in MgB$_2$ was demonstrated for 
the magnetic field parallel to the $c$-axes of the crystal and also for the field in $ab$-plane 
and both of which show anomalous behavior\cite{matsuda}. 
We use the parameters $H_{c2}^{(2)} / H_{c2}^{(1)} = 0.04$ 
and $\rho_n^{(2)}/\rho_n^{(1)}=0.26$ for $H \parallel ab$ case, and $H_{c2}^{(2)} / H_{c2}^{(1)} =0.06$ 
and $\rho_n^{(2)}/\rho_n^{(1)}=0.5$ for $H \parallel c$ case.
The order of these values 
appear to be consistent with the facts that the DOS for each two bands are roughly equal, 
the energy size of larger gap is 7 mev and that of smaller gap is 2mev 
\cite{review}, and the carrier scattering rates for each 
two bands are almost equal\cite{matsuda}.  
As is given in Fig. \ref{rho2_f}, our result agrees well    
with experimental data, especially for $H \parallel c$ case.

\begin{figure}[h]
\vspace{0.5cm}
\centerline{
\epsfysize8cm\epsffile{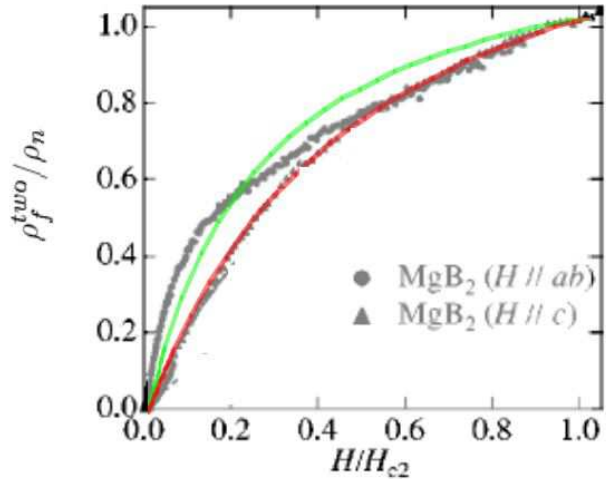}}
    \caption{ We compare the experimental 
data of flux flow resistivity in MgB$_2$ given by Ref. \cite{matsuda} and 
our result Eq. (\ref{field-vs-rho-f-two}). 
Black dots and triangles show the experimental results for $H \parallel ab$ case 
and $H \parallel c$ case, respectively. Green line
shows Eq. (\ref{field-vs-rho-f-two})  with parameters 
$H_{c2}^{(2)} / H_{c2}^{(1)} = 0.04$ and $\rho_n^{(2)}/\rho_n^{(1)}=0.26$, and  
Red line shows that with $H_{c2}^{(2)} / H_{c2}^{(1)} =0.06$ 
and $\rho_n^{(2)}/\rho_n^{(1)}=0.5$.} 
\label{rho2_f}
\vspace{0.5cm}
\end{figure}

In summary, we have considered a system with two superconducting bands. 
The total flux flow resistivity in the system 
is written as the parallel connection of the flux flow resistivity 
for the 1st-band vortices and that for the 2nd-band vortices. 
The result agrees with anomalous field dependence of 
the flow resistivity recently observed in the 
two-gap superconductor MgB$_2$.
Our discussion would be generalized to 
the system with $n(>2)$ superconducting bands. 
The detailed discussion will be published elsewhere. 

%
%
%
%

%
%
%



\begin{thebibliography}{00}


\bibitem{Tinkham} ``Introduction to Superconductivity'', 
M. Tinkham (2nd. ed., McGraw-Hill, Singapore, 1996), and references therein. 



\bibitem{matsuda1} Y. Matsuda, {\it et al} Phys. Rev. B {\bf 66} (2002) 014527. 


\bibitem{nagamatsu} Nagamatsu, N. {\it et al.} Nature (London) 
\textbf{410} (2001) 63.


\bibitem{review} See, for the recent review, P. C. Canfield and G. W Crabtree, 
Physics Today {\bf 56} (3), (2003) 34. 







\bibitem{matsuda} A. Shibata {\it et al.} 
Phys. Rev. B {\bf 68} (2003) 060501(R). 





\end{thebibliography}
\end{document}